\documentclass[twocolumn,english,preprintnumbers,aps,prd,floatfix,showpacs,amsmath,amssymb]{revtex4}

\usepackage{graphicx}
\usepackage{epsfig}
\usepackage{latexsym}
\usepackage{amssymb}

\newcommand{\tmass}{\mbox{$\tilde M$}}

\begin{document}
\title{Dark energy constraints from lensing-detected galaxy clusters}
\author{Laura Marian and Gary M. Bernstein} 
\affiliation{Department of Physics and Astronomy, University of Pennsylvania, Philadelphia, Pennsylvania 19104, USA}

\date{January 26, 2006}

\begin{abstract}

We study the ability of weak lensing surveys to detect galaxy clusters
and constrain cosmological parameters, in particular the equation of state of
dark energy. 
There are two major sources of noise
for weak lensing cluster measurements: the ``shape noise'' from the
intrinsic ellipticities of galaxies; and the large scale projection
noise.  We produce a filter for the shear field
which optimizes the signal-to-noise of shape-noise-dominated
shear measurements. Our Fisher-matrix analysis of this projected-mass
observable makes use of the shape of this mass-function, and takes
into account the Poisson variance, sample variance, shape noise, and
projected-mass noise, and also the fact that the conversion of the
shear signal into mass is cosmology-dependent.  The Fisher analysis is
applied to both a nominal 15,000~deg$^2$ ground-based survey and a
1000~deg$^2$ space-based survey.
Assuming a detection threshold of $S/N=5$, we find both experiments
detect $\approx20,000$ clusters, and yield 1-$\sigma$ constraints
of $\Delta w_{0}\approx0.07$, $\Delta w_{a}\approx0.20$ when combined
with CMB data (for flat universe).  The
projection noise exceeds the shape noise only for
clusters at $z\lesssim0.1$ and has little effect on the derived
dark-energy constraints.  Sample variance does not significantly
affect either survey.
Finally, we note that all these results are extremely sensitive to the
noise levels and detection thresholds that we impose. They can be
significantly improved if we combine ground and space surveys as
independent experiments and add their corresponding Fisher matrices.
 \end{abstract}

\maketitle
\section{Introduction}

\par Constraining the dark energy equation of state and density
parameters is the objective of many cosmologists; not few of them have
considered using present or future cluster data to attain this goal.
Cluster methods rely mostly on detection and counting of objects using
some mass-indicating observable; for constraining cosmology there are
4 ways of finding clusters: optical emission by galaxies, the X-ray
emission by the hot
intracluster medium, the Sunyaev-Zeldovich effect on CMB, and
gravitational lensing.

\par X-ray clusters received a lot of attention from cosmologists 
\cite{HA5,H2,WS1,KS7,BF8,LS12}, then SZ clusters
\cite{EC6,HH10,WB13,BW14}.  See \cite{HM11,MM15,MM16} for thorough
comparisons between the efficiency of 
extracting cosmological information from X-ray and SZ clusters and
also for using complementary studies of the CMB and supernovae
distance measurements to improve constraints.
For optically-detected cluster surveys and their results see the work of
 \cite{Opt1,Opt2,Opt3,Opt4,Opt5}. 
\par The main issue with the X-ray, optical, and SZ clusters is the so-called
mass-observable relation: cluster masses are not measured directly,
but they are estimated from the real observables, such as the X-ray
temperature or flux or the SZ integrated flux. But, as pointed out in
\citep{MM15}, the mass-observable relation can have non-standard
redshift evolution and, if not carefully calibrated, the constraints
on dark energy parameters are compromised. Cross-calibration of the
mass-observable relations between different types of surveys could,
though perhaps not very efficiently, ameliorate the situation; more
recently, self-calibration using the sample variance of the counts due
to clustering of clusters has been proposed as a non-costly
alternative to cross-calibration, eg see \cite{LH1,HK3}.

\par Weak gravitational lensing detects clusters via the slight
distortions imparted on the images of background galaxies.  The
lensing information is obtained from shear maps: clusters are detected
using some filtering technique that finds the points where the
signal-to-noise ($S/N$) is high enough for us to conclude that there is
an overdensity.  The filtered shear is the mass observable. The
relation of this observable to the projected mass is very simple and
unambiguous.  The strength of the lensing method is this lack of
ambiguity in the mass-observable relation.

A difficulty arises because such a point with high $S/N$ does not
necessarily correspond to a virialized cluster: it could also be the
result of unvirialized large scale structure projected along the line
of sight, or the superposition of multiple unrelated lower-mass
objects \cite{HS4,Takada}.  Even for virialized clusters, there will be
substantial scatter between the projected-mass observable and the
traditionally defined virial mass \cite{CM}.  But there is in fact
no underlying need for lensing observables to correspond to other
cluster observables, nor even to a dynamical definition of mass.  We
will argue that counts of projected-mass overdensities are just as
useful for cosmology as counts of virialized halos.

\par In this paper we study how well a $4-$dimensional parameter space
can be constrained using weak-lensing-detected clusters.  The
parameters in question are $\Omega_m$ (the matter density parameter), $w_0, w_a$and $\sigma_8$ (the amplitude of the matter power spectrum). $w_0$ and $w_a$ define the time-varying equation of state of dark energy: $w(a)=w_0+(1-a)w_a$.  We use as
examples two proposed future surveys, a ground-based Large Survey
Telescope (LST) \cite{lst} and the space-based Supernova Acceleration Probe
(SNAP) \cite{snap}. We first review the role
of cluster-mass observables in constraining cosmology.  In
\S\ref{snmax} we describe a filter based on $S/N$ maximization of
shear measurements that determines the minimum detectable mass of an
object placed at a certain redshift.  In \S\ref{eonly} we calculate
the cosmological constraints obtained when assuming that the intrinsic
ellipticities of background galaxies are the only source of noise.  In
\S\ref{projection} we treat the large scale structure projection
errors and in \S\ref{conc} we draw some conclusions.

\par The fiducial $\Lambda$CDM cosmological model for this whole paper
is: flat universe, $\Omega_{m}=0.27$, $\sigma_{8}=0.9$, $w_{0}=-1$, $w_{a}=0$,
$h=0.72$, $\Omega_{b}\,h^{2}=0.024$, consistent with the first-year
{\it WMAP} results \cite{WMAP}.

\section{Mass Observables and Cosmology}
\label{Gary} Any observable statistic can be used as a cosmological
test, as long as (1) it can be measured on the real Universe, and (2)
its value can be predicted as a function of the cosmological
parameters of interest.  The utility of the statistic depends upon:
the accuracy with which the measurement can be made; the accuracy with
which the {\em predictions} can be made; and the sensitivity of the
statistic to the parameters.  To forecast the parameter accuracies, as
we aim to do here, we need only estimate these three characteristics
of the statistic.

Overdensity-counting methods have as their statistic the distribution
$dN/d\Omega\,dz\,d\tmass$ of ``clusters'' in a solid angle $\Omega$ vs redshift as a function of
some observable quantity \tmass\ [the clustering of these
overdensities may be an additional statistic].  In an ideal world, the
observable would be the virial mass $M$, because there exist analytic
frameworks \cite{PS,ST10} for predicting their
distribution, a.k.a. the mass function.  To constrain dark energy at
interesting levels, the mass function must be predicted to an accuracy
that will undoubtedly require $N$-body simulations, not just the
analytic frameworks. For more recent work on mass functions and their accuracy see for instance \cite{Mf1,Mf2}.  This is also of course true for the real-life
substitutes for virial mass: the x-ray flux/temperature, the SZ
decrement, galaxy counts, and the lensing shear.  There is no
long-term advantage to 
an observable that is closely correlated to the virial mass.  The
ultimate utility of these methods will depend upon the fidelity of the
numerical predictions, and here the lensing method is clearly
superior.  The shear prediction needs only the mass distribution, and
83\% of the mass is easily-modelled collisionless dark matter.
X-rays and SZ decrements depend fully upon the more complex baryon
distribution, and the electron temperature as well.  Cooling and
density fluctuations particularly affect the x-ray predictions.
Galaxy counts are even more difficult to predict.  It would be bold to
assert that modelling will ever predict any of these observables {\em
other} than the shear to the percent-level accuracies we will someday
desire.

Are analytic mass functions adequate for forecasting parameter
accuracies?  To be so, they must roughly---but not exactly---predict
the number of peaks in \tmass, so that our Poisson errors are properly
estimated.  When \tmass\ is a projected mass measurement, several
numerical studies \cite{HS4,Takada} show that up to tens
of percent of detections can be ``false positives'' in the sense of
having no corresponding virialized cluster, and some virialized
clusters are missed.  The Poisson statistics are grossly
perturbed only when the $S/N$ threshold is low enough that
measurement-noise peaks overwhelm the mass signals.

The analytic model must also properly capture the dependence upon
cosmological parameters.  The ``false positives'' when \tmass\ is the
projected mass are not virialized, but they are real structures whose
abundance will scale with 
$\sigma_8$ and the linear growth rate in a manner not grossly
different from the virialized structures \cite{WK6}. 

The projections which distinguish lensing-derived projected masses
from dynamical masses can be divided into two classes: first, there
are projections between mass structures that are widely
separated along the line of sight, in which case there is no angular
correlation between the projected halos.  In \S\ref{projection} we
treat such projections as a source of random noise on the mass
determinations of detected clusters induced by projection of
below-threshold halos. 

The second difference between lensing mass and virial mass is that the
former includes all the structure along the line of sight that is
correlated with the mass peak, not just the virialized (or
unvirialized) core.  The lensing ``mass function'' is therefore
distinct from the mass function of virialized halos, but it is just as
well defined.  We will simply assume for this paper that the
projected-mass function is equal to
the virial-mass function produced by the Press-Schechter formalism and
related techniques, in the absence of numerical or analytic evidence
that the two differ dramatically.

We conclude that the distinction between the weak-lensing
projected-mass observable \tmass\ and the virial mass $M$ will not be
a barrier to its use as a precision cosmological constraint, and that
we can use the virial mass functions for approximate forecasts of
these constraints. Naturally, for accurate constraints on dark energy, one will have to use a lensing mass function derived from numerical simulations.   

\section{The minimum detectable mass}
\label{snmax}

\par Given a cluster at some redshift, we would like to determine the
smallest value of its mass such that it could be detected through
weak lensing (WL) effects.  In this section we assume that the
intrinsic ellipticities of the galaxies represent the dominant source
of noise for the projected mass measurement. We shall test this
assumption in \S\ref{projection}, where we consider the large scale
projection effects.

\subsection{S/N maximization}

\par As mentioned in the introduction, for WL measurements the
observable is the shear, which encodes information about the direction
and magnitude of the distortion of background galaxies images. The
components of the shear contain derivatives of the deflection angle
with respect to the apparent position of the source galaxy.  In the
case of WL, distortions and magnifications are so small that we can
very well approximate the relation between the components of the induced shear
$(\gamma_{1},\gamma_{2})$ and those of the measured image ellipticity
$(e_{1},e_{2})$ in the following way (see \cite{ModCos} for example):
\begin{equation} 
e_{i}\approx
2\gamma_{i}+\tilde{e}_{i}\,, \qquad i\in\{1,2\}\;. 
\label{eq:e} 
\end{equation}
$(\tilde{e}_{1},\tilde{e}_{2})$ represents the intrinsic ellipticity
of the galaxy. From
measurements of $N$ galaxies, the shear can be estimated with the
accuracy:
\begin{equation} 
{\rm Var}(\gamma_{1})=\frac{\sigma_{\gamma}^{2}}{N}\,,
\label{eq:vargamma}
\end{equation} 
where $\sigma_{\gamma}$ is the uncertainty in the measurement of one galaxy. For a detailed investigation of the shape noise, we refer the reader to \cite{BJ02}. An approximate expression is \cite{G}:
\begin{equation} 
\sigma_{\gamma}\approx\frac{\langle
e_{1}^{2}\rangle^{1/2}}{2}\,. 
\label{eq:ellip} 
\end{equation}
\par A filtered shear map is used to detect clusters by selecting the
points where the $S/N$ peaks above some threshold, $(S/N)_{\rm{min}}$.
But the filtered shear values are also used in the interpretation of
the detected clusters.  We are at liberty to choose any filtering
method we wish, provided we can tie its results to a physical quantity
for which there exists a theory. 
We choose to normalize our filter to reproduce the virial mass when
applied to canonical clusters, namely spherically
symmetric objects, with NFW density profiles in the context of a
$\Lambda$CDM cosmology---i.e. what we expect the average (if not
typical) cluster profiles to be.
This close scaling will allow us to use the well-studied and
tested mass function theory to predict the number of detectable
clusters. Therefore, the two requirements for our filter are first
that it yield a value closely related to the virial mass for canonical
clusters; and second, that the virial mass estimator produce a maximum
$S/N$ from the shear map.  

For a circularly symmetric lens, the shear has only one component,
tangent to the annulus about the cluster center.
We define our mass estimator as a weighted sum of the shears in such
annuli:
\begin{equation}
\tmass=\sum_{k,i}w(\theta_{k},z_{i})\gamma_{T}(\theta_{k},z_{i})\,, 
\label{eq:Mtilda} 
\end{equation}
where $k$ designates the annulus with angular radius $\theta_{k}$,
$\gamma_{T}(\theta_{k},z_{i})$ is the measured shear of annulus $k$ on
sources in redshift bin $i$, and $w(\theta_{k},z_{i})$ is the weight
on this shear.  The variance of this estimator is:
\begin{equation}
{\rm Var}(\tmass)\!=\!\!\!\sum_{k,k',i,j}\!\!w(\theta_{k},z_{i})w(\theta_{k'},z_{j}){\rm
  Cov}\left[\gamma_{T}(\theta_{k},z_{i})\gamma_{T}(\theta_{k'},z_{j})\right]. 
\label{eq:FullVar}
\end{equation}
When the noise of the intrinsic ellipticities of
galaxies is dominant, the contributions of different redshift bins to
the estimator variance are uncorrelated and the above expression
reduces to:
\begin{equation} 
{\rm Var}(\tmass)=\sum_{k,i}w(\theta_{k},z_{i})^{2}\
\,{\rm Var}(\gamma_{T}(\theta_{k},z_{i}))\,. \label{eq:noise1}
\end{equation}
The desired maximization of $S/N$ is obtained for the following set of
$w$'s:
\begin{equation}
w(\theta_{k},z_{i})\propto\frac{\gamma_{T}(\theta_{k},z_{i})}{{\rm
Var}(\gamma_{T}(\theta_{k},z_{i}))},\; \forall\,
k, i\,. 
\label{eq:w} 
\end{equation}
The normalization constant in
equation~(\ref{eq:w}) is set by the first property of our filter: if
the real clusters have NFW profiles and the real cosmology is
$\Lambda$CDM, then the estimator must return the virial mass of the
clusters. Thus, the constant is:
\begin{equation}
\mathcal{C}=\frac{M_{\rm{vir}}}{\sum_{k,i}(\gamma_{T}^{\Lambda
CDM}(\theta_{k},z_{i}))^2/ \rm{Var}(\gamma_{T}(\theta_{k},z_{i}))}\,.
\label{eq:constant}
\end{equation}
The shear and the optimal weight
separate into a part depending only on the lens properties,
and a part depending only on the lens and source redshifts. For
redshift bin $i$ we write:
\begin{equation}
\gamma_{T}(z_{i},z_{d})=\frac{\int_{z_{i}}^{z_{i+1}}dz_{s}\,\mathcal{P}(z_{s})Z(z_{s},z_{d})\,\gamma_{\infty}(z_{d})}{\int_{z_{i}}^{z_{i+1}}dz_{s}\,\mathcal{P}(z_{s})}\,.
\label{eq:Z1}
\end{equation} 
In the above equation, $\gamma_{\infty}(z_{d})$ is the
shear of a hypothetical source at infinity, $\mathcal{P}(z_{s})$ is
the redshift distribution of source galaxies (see \S\ref{details}) and
$Z$ is given by:
\begin{equation} 
Z(z_{s},z_{d})=\left\{\begin{array}{ll}
\frac{D_{ds}}{D_{s}}& \mbox{if $z_{s}>z_{d}$}\\ 0& \mbox{otherwise}
                       \end{array}\right.\,,
\end{equation} 
where $D_{ds}$ is the angular-diameter distance between
the lens and the source and $D_{s}$ is the angular-diameter distance between theobserver and the source.  Throughout this paper we consider only the
case of a flat universe.  In the limit of an infinite number of
redshift bins and if we introduce the shear per unit mass for the
canonical cluster, $\tilde\gamma_\infty$, equation~(\ref{eq:Z1})
becomes:
\begin{equation} 
\gamma_{T}(z_{i},z_{d},\theta)=\tmass
Z(z_{i},z_{d})\,\tilde{\gamma}_{\infty}(z_{d},\theta)\,.
\label{eq:Z2}
\end{equation}
To make the notation easier, we shall ignore the dependence on
$z_{d}$ of $\gamma_{T}$ and $\tilde{\gamma}_{\infty}$.  Using the
normalization constant defined by equation~(\ref{eq:constant}), we
obtain for the noise of the measurement:
\begin{eqnarray} 
{\rm Var}(\tmass)=
\left(\sum_{k,i}\frac{\tilde\gamma_{\infty}(\theta_{k})^{2}Z(z_{i},z_{d})^2}{{\rm Var}(\gamma_{T}(\theta_{k},z_{i}))}\right)^{-1}\,,
\label{eq:noise2}
\end{eqnarray} 
with $\tilde \gamma_{\infty}$ and $Z$ evaluated for the
NFW profile and a $\Lambda$CDM cosmology. ${\rm Var}(\gamma_{T}(\theta_{k},z_{i}))$ is given by
equation~(\ref{eq:vargamma}):
$${\rm Var}(\gamma_{T}(\theta_{k},z_{i}))=\frac{\sigma_{\gamma}^{2}}{N(\theta_{k},z_{i})}\,,$$
where $N(\theta_{k},z_{i})$ is the number of galaxies in redshift bin
$i$ sheared in annulus $k$ and
$$N(\theta_{k},z_{i})=\int_{z_{i}}^{z_{i+1}}\int_{0}^{2\pi}\int_{\theta_{k}}^{\theta_{k+1}}dz_{s}d\varphi\,d^{2}\theta\mathcal{P}(z_{s})n(\theta,\varphi)\,.$$
If the annuli are dense enough and using the above relations, as well
as equations~(\ref{eq:w}) and~(\ref{eq:constant}), we can rewrite the
mass estimator of an NFW cluster as:
\begin{equation} \tmass=M_{\rm{vir}}\,g({p_{\alpha}})\,,
\label{eq:Mtilda2}
\end{equation} 
where the $p_{\alpha}$'s generically denote
cosmological parameters. $g({p_{\alpha}})$ is the change of the
estimated mass from the true virial mass if the cosmology is different
from $\Lambda$CDM:
\begin{equation}
g(p_{\alpha})=\frac{\int_{0}^{\infty}dz_{s}\mathcal{P}(z_{s})\int_{0}^{\theta_{lim}}
d\theta\,\theta\tilde{\gamma}_{T}(\theta, z_{s}, z_{d})
\tilde{\gamma}_{T}^{m}(\theta, z_{s}, z_{d})}{\int_{0}^{\infty}dz_{s}\mathcal{P}(z_{s})\int_{0}^{\theta_{lim}}
d\theta\,\theta\tilde{\gamma}_{T}^{2}(\theta, z_{s}, z_{d})}.
\end{equation} 
 $\tilde \gamma_{T}(\theta, z_{s}, z_{d})$ is the tangential shear per
unit mass for a $\Lambda$CDM cosmology. $\tilde \gamma_{T}^{m}$
is the shear one will measure in the real cosmology,
possibly other than $\Lambda$CDM.

Note that if $\gamma_\infty(\theta)$ is independent of cosmology, then
$g(p_\alpha)$ is simply a ratio of angular-diameter distances.  More
generally, the parameters of the NFW profile of a cluster of given
$M_{\rm vir}$ will depend upon cosmology.

The variance of the mass estimator,
written also in the case of continuous annuli is:
\begin{eqnarray} 
\lefteqn{{\rm Var}(\tmass,z_{d})=
\frac{\sigma_{\gamma}^{2}}{2\pi n}} & \nonumber \\
\nonumber\\
&\times\left(\int_{0}^{\infty}dz_{s}\mathcal{P}(z_{s})Z^{2}(z_{s},z_{d})
\int_{0}^{\theta_{lim}}d\theta\,\theta\,
\tilde{\gamma}_{\infty}^{2}(\theta,z_{d})\right)^{-1}.
& \label{eq:noise3}
\end{eqnarray} 
Here we have considered a constant angular
concentration of galaxies. For the upper bound of the second integral
in the above equation we take $\theta_{lim}=\frac{2\,R_{vir}}{D_{d}}$,
with $R_{vir}$ the virial radius of the measured cluster.  This choice
of $\theta_{lim}$ seems a reasonable trade-off for looking widely
enough to get a significant signal without too much contamination by
other lensing structures.

A cluster is detectable if its mass estimator has a minimum value of
\begin{equation} 
\tilde M_{\rm min}(z_{d})=(S/N)_{\rm min}\:\sqrt{{\rm Var}(\tmass,z_{d})}\,,
\label{eq:Mmin}
\end{equation} 
where ${(S/N)_{\rm min}}$ is the detection threshold
that we choose for our measurement and ${\rm Var}(\tmass,z_{d})$ is
given by equation~(\ref{eq:noise3}).

Let us emphasize again that only in the case of a canonical cluster
(NFW profile in a $\Lambda$CDM cosmology) does our estimator translate
the shear signal into the cluster's virial mass. If the cosmology is
changed from $\Lambda$CDM, the same shear signal results from a
different virial mass, as one can see from
equation~(\ref{eq:Mtilda2}). But even so we are still able to find and
count objects. Since $N$-body simulations will yield the functional
dependence on cosmology of 
the difference between the estimated and the virial mass, we can still
understand our counts and know how to extract cosmological constraints
from them.  The measurement is well-defined for clusters that do not
have the canonical NFW profile, but in this paper we do not attempt 
numerical simulations to derive the effect of this difference on WL
cluster counts. 

\subsection{Distribution of sources and the NFW shear}
\label{details}

\par We now present a few steps that we take in order to compute the
variance given by equation~(\ref{eq:noise3}).
\par We assume the source galaxies to have the redshift distribution
introduced by \cite{Smail} and used in many other works, e.g. see
\cite{BS9,WK6,HS4}:
\begin{equation}
\mathcal{P}(z)=\frac{1}{2\,z_{0}^{3}}\,z^{2}\,\exp{(-z/z_{0})}\,. \label{eq:sourcedistrib} 
\end{equation}
The mean of this distribution is $\langle z \rangle=3 z_{0}$ ; the
characteristics of the fiducial survey will set the value of $z_{0}$.

As already specified, we optimize our filter for spherically symmetric
clusters with NFW profiles and in a $\Lambda$CDM cosmology.
\par A thorough treatment of NFW lenses is given by \cite{B8}, as
well as by \cite{WB7}. The shear due to NFW mass distributions is:
\begin{equation} 
\gamma_{\rm NFW}(x)= \left\{ \begin{array}{lcl}
\frac{r_{s}\,\delta_{c}\,\rho_{c}}{\Sigma_{crit}}\,g_{<}(x)&\mbox{$,\;$
$x<1$ }\\
\frac{r_{s}\,\delta_{c}\,\rho_{c}}{\Sigma_{crit}}\,\left[10/3 +
4\ln(1/2)\right]&\mbox{$,\;$ $x=1$ }\\
\frac{r_{s}\,\delta_{c}\,\rho_{c}}{\Sigma_{crit}}\,g_{>}(x)
&\mbox{$,\;$ $x>1$}\,, \end{array}\right.
\end{equation} 
where $x=\frac{\theta D_{d}}{r_{s}}$ and $r_{s}$ is the
scale radius of the halo. $\delta_{c}$ is the characteristic
overdensity and $\rho_{c}(z)=\frac{3\,H^{2}(z)}{8\pi G}$ is the
critical density of the universe. The characteristic overdensity is related to the concentration parameter of halos  by the condition that the mean density within the virial radius $R_{vir}$ should be $\Delta_{vir}\rho_{c}$, where $\Delta_{vir}=200$ for NFW: $\,\delta_{c}=\frac{\Delta_{vir}}{3}\,\frac{c^{3}}{\ln (1+c)-\frac{c}{1+c}}$ (see \cite{NFW9} for more details). The functions $g_{<}$ and $g_{>}$
are independent of cosmology and of cluster parameters and are given
by:
$$\begin{array}{lc}

                   g_{<}(x)=
\frac{8\tanh^{-1}\sqrt\frac{1-x}{1+x}}{x^{2}\sqrt{1-x^2}}
+\frac{4}{x^{2}}\ln\left(\frac{x}{2}\right) -\frac{2}{x^{2}-1}\\
\mbox{$\;\;\;\;$}\mbox{$\;\;\;\;$}\mbox{$\;\;\;\;\;$}
+\frac{4\tanh^{-1}\sqrt\frac{1-x}{1+x}}{(x^{2}-1)\sqrt{1-x^{2}}}\,;

            \end{array}$$

$$\begin{array}{lc}

                   g_{>}(x)=
\frac{8\tan^{-1}\sqrt\frac{x-1}{x+1}}{x^{2}\sqrt{x^{2}-1}}
+\frac{4}{x^{2}}\ln\left(\frac{x}{2}\right) -\frac{2}{x^{2}-1}\\
\mbox{$\;\;\;\;$}\mbox{$\;\;\;\;$}\mbox{$\;\;\;\;\;$}
+\frac{4\tan^{-1}\sqrt\frac{x-1}{x+1}}{(x^{2}-1)^{3/2}}\,.

             \end{array}$$
Given a cosmology and a halo of some mass and at some redshift, we
need the concentration parameter $c$ in order to compute the
characteristic overdensity and the scale radius. We assume a relation
between $c$ and $M$ as given by \cite{NFW9}.

\subsection{Mass thresholds for nominal surveys}

\par
\begin{figure}[h!]  \centering
\includegraphics[scale=0.8]{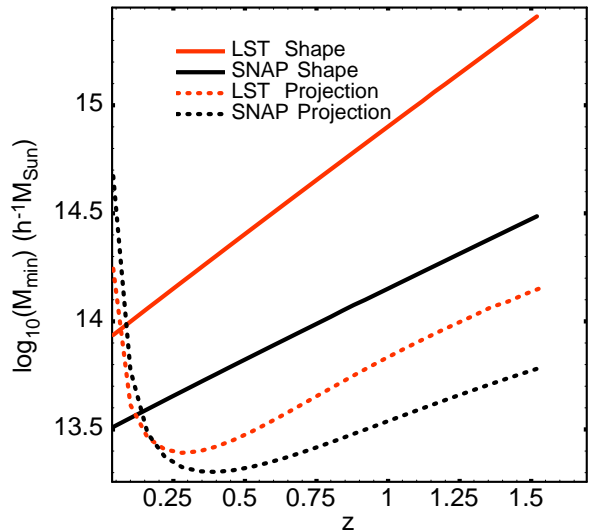}
\caption{Minimum detectable mass for LST and SNAP when
$\rm{(S/N)_{min}}=5$. The solid lines represent $M_{\rm{min}}$ when
the intrinsic ellipticity noise dominates the measurement. The dotted
lines represent $M_{\rm{min}}$ when the projection noise is dominant.}
\label{fig:Mcrit5}
\end{figure}

 In order to see the redshift dependence of the WL mass thresholds
provided by equations~(\ref{eq:noise3}) and~(\ref{eq:Mmin}), we
consider the examples of possible two future surveys: a Large Survey
Telescope (LST) from the ground, and the space-based {\it Supernova
  Acceleration Probe (SNAP)}.
We assume for both surveys $\sigma_{\gamma}=0.3$.  For LST we use a
distribution of galaxies with a mean redshift of 1,
i.e. $z_{0}=0.33$. The angular concentration of sources is $n=30$
galaxies/arcmin$^{2}$ and the survey area is
$\mathcal{A}=15000$deg$^{2}$. In the case of SNAP, we take
$z_{0}=0.5$, $n=100$ galaxies/arcmin$^{2}$ and
$\mathcal{A}=1000$deg$^{2}$.  For a fixed $z_{d}$ we compute ${\rm
Var}(\tmass,z_{d})$ as given by equation~(\ref{eq:noise3}) and then we
find the smallest value of $\tmass$ verifying ~(\ref{eq:Mmin}). This
value gives $\tmass_{\rm min}(z_{d})$. The solid lines in
figure~\ref{fig:Mcrit5} show the detection thresholds for both
instruments using the NFW definition of the virial mass. To compute
the cluster abundance, we shall convert to the Sheth-Tormen definition
of virial mass and $\tmass_{\rm min}$ will rise by approximately 30\%
for $z_{d}<0.4$.

Not surprisingly, the detection threshold for the space telescope is
much lower than that of the ground telescope (3 or 4 times for small
redshifts and almost an order of magnitude for redshifts higher than
1). In compensation, LST covers a substantially larger area of the
sky, so in the end both telescopes can detect similar abundances of
clusters.

Using equation~(\ref{eq:nc}), we obtain $\approx 21500$
detectable clusters for LST and $\approx 17500$ for SNAP if
$(S/N)_{\rm{min}}=5$. If $(S/N)_{\rm{min}}=10$, the numbers are 1600
and 2200 respectively.  As the cluster mass function is very steep,
the derived cluster counts and resultant cosmological constraints are
quite sensitive to assumptions about input noise levels
($\sigma_\gamma$, $n$, and $z_d$) and $(S/N)_{\rm min}$.

 \begin{figure}[h!]  \centering
 \includegraphics[scale=0.82]{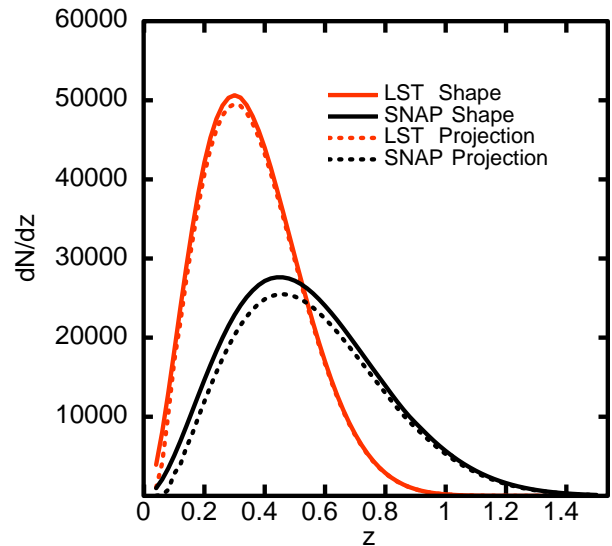}
\caption{Cluster redshift distribution for LST and SNAP when
$\rm{(S/N)_{min}}=5$. The solid lines correspond to the case when the
shape noise dominates. The dotted lines show the cluster distribution
when in addition to the shape noise we also include projection effects
in the minimum detectable mass.}
\label{fig:ncANs5}
\end{figure}

\section{Fisher matrix calculations}
\label{eonly}

\par In this section we use the Fisher information matrix to estimate
how well WL cluster surveys can constrain the following parameters:
$\Omega_{m}$, $\sigma_{8}$, $w_{0}$ and $w_{a}$. First we compute the
number of clusters detectable by our fiducial surveys. Then we
calculate the Fisher matrix considering Poisson noise and sample
variance noise.  Throughout this section we hold to the assumption
that the intrinsic ellipticities of galaxies are the main
noise source in the cluster measurements.

\subsection{Cluster abundances}
 \par As mentioned in section \S\ref{Gary}, the identification of the
projected mass with the virial mass allows us to calculate the number
of detectable clusters as an integral of the mass function$\,\frac{dN}{dV\,dM}$.  The
number of clusters per unit solid angle is
\begin{equation}
\frac{dN}{d\Omega}=\int_{0}^{\infty}dz\frac{dV}{dz\,d\Omega}\,\int_{M_{{\rm
min}}(z)}^{\infty}d\ln{M}\,\frac{dN(M,z)}{dVd\ln{M}}\,. \label{eq:nc}
\end{equation} 
Here dV is the comoving volume element and
$\rm{M_{min}(z)}$ is the virial mass of a cluster still detectable at
redshift $z$, obtained from equations~(\ref{eq:Mtilda2})
and~(\ref{eq:Mmin}) from section \S\ref{snmax}.  The comoving number
density of halos is given by:
$\frac{dN}{dVd\ln{M}}\,d\ln{M}=\frac{\rho}{M}f(\nu)d\nu$, where $\rho$
is the comoving matter density and $f(\nu)$ is a semi-analytic dimensionless form of the mass function.
$\nu=\frac{\delta_{sc}^{2}}{\sigma^{2}}$, $\delta_{sc}$ is the
critical density for spherical collapse and $\sigma^{2}(M,z)$ is the
linear density field variance, smoothed with a top-hat filter.
$\sigma(M,z)$ is the only quantity in the mass function that depends
on the dark energy equation of state through the growth
factor: 
$$\sigma(M,z)=D(z)\,\sigma(M,z=0)\,.$$ 
We assume a Sheth-Tormen $f(\nu)$ \cite{ST10}:
$$f_{ST}(\nu)=A \sqrt\frac{a}{2\pi\nu}\left[1+(a\nu)^{-p}\right]\exp(-a\nu/2)\,,$$
with $A=0.3222, a=0.707, p=0.3, \delta_{sc}=1.69$.
\subsection{The Fisher matrix}

The Fisher matrix calculations were completed for cells corresponding
to pairs of bins in redshift and filtered shear $\{z,\tmass \}$. This
choice of binning has two major benefits. Firstly, we use a directly measured
quantity, the (filtered) shear \tmass, rather than an inferred mass, to
describe the data.  Thus we can monitor
how both the number of objects in a bin and their masses change with
cosmology, in accordance with equations~(\ref{eq:Mtilda2})
and~(\ref{eq:nc}). Secondly, we retain information on the number vs
\tmass\ 
i.e. we use the shape of the mass function. Most papers
dealing with cluster Fisher matrix estimations wash away the
information given by the shape of the mass function by integrating the
mass function over mass above threshold. We considered 50
$\tmass$-bins and 35 redshift bins, going from 0 to 2.

Ignoring the errors in redshift determinations (for both galaxies and
clusters), bin counts are affected by 2 kinds of noise: Poisson noise
and sample variance noise. The latter arises from the fact that the
large scale structure of the universe correlates the number density of
virialized objects in different volume elements.  \cite{HK3}
have studied the importance of sample variance noise relative to
Poisson noise in cluster surveys . They concluded that sample variance
can dominate Poisson noise for low-mass clusters,
$M\leq\,4\times10^{14}M_{\odot}/h$.
\par The Poisson Fisher matrix is defined as:
\begin{equation} 
F_{\alpha\beta}=\sum_{i=1}^{I}
N_{i,\alpha}\,N_{i,\beta}\frac{1}{N_{i}}\,,
\label{eq:fishPoisson}
\end{equation} 
where $N_{i,\alpha}=\partial N_{i}/\partial p_{\alpha}$
and $p_{\alpha},p_{\beta}$ can be any of the four parameters mentioned
earlier. $N_{i}$ is the estimated number of clusters for bin $i$ and
$I$ is the total number of bins.
\par To account for the correlations between bins, we follow
\cite{LH1} and employ an approximation of the Fisher matrix from the
limiting cases of Poisson dominance and sample variance dominance:
\begin{equation}
F_{\alpha\beta}=\rm{N_{,\alpha}^{t}\,C^{-1}N_{,\beta}}+\frac{1}{2}\,\rm{Tr}[C^{-1}\,S_{,\alpha}\,C^{-1}\,S_{,\beta}]\,. 
\label{eq:fishPSV}
\end{equation} 
$\rm{N}$ is a vector of length $I$ whose elements are
the $N_{i}$'s from above.  $C$ is the covariance matrix (dimension
$I\times I$), a sum of the shot noise and the sample covariance noise:
$C=S+\mathcal{N}$. $\mathcal{N}$ is a diagonal matrix with $N_{i}$ on
its diagonal.  The sample covariance matrix is (e.g. \cite{HK3}):
$$S_{ij}=b_{i}b_{j}N_{i}N_{j}D_{i}D_{j}\int\frac{d^{3}k}{(2\pi)^{3}}W_{i}(\vec{k})W_{j}^{*}(\vec{k})P_{\rm{lin}}(k)\,.$$
Here $i$ and $j$ are bin indices, $P_{\rm{lin}}(k)$ is the linear
power spectrum and $b$ is the linear bias as given by
\cite{ST10} :
$$b(M,z)=1+\frac{a\delta_{sc}^{2}/\sigma^{2}-1}{\delta_{sc}}+\frac{2p}{\delta_{sc}[1+(a\delta_{sc}^{2}/\sigma^{2})^p]}\,,$$
where $a$, $p$ and $\delta_{sc}$ have the same values as for the mass
function. $D$ is the linear growth factor and $W_{i}(\vec{k})$ are
the survey windows. As suggested in \citep{HK3}, we considered the
survey windows a series of concentric cylinders (redshift slices) at
comoving distance $r_{i}$, of height $\delta r_{i}$ $(\delta r_{i}\ll
r_{i})$, all subtending the same angle $\theta_{s}$.  In Fourier space
such windows have the expression:
$$W_{i}(\vec{k})=2\exp({ik_{\parallel}r_{i}})\,\frac{\sin({k_{\parallel}}\delta r_{i}/2)}{k_{\parallel}\delta r_{i}/2}\,\frac{J_{1}(k_{\perp}r_{i}\theta_{s})}{k_{\perp}r_{i}\theta_{s}}\,,$$
where $k_{\parallel}$ and $k_{\perp}$ are the components of $\vec{k}$,
parallel and perpendicular to the line of sight and $J_{1}$ is the
first order Bessel function.
\subsection{Results}
\label{res} To the cluster constraints we have added CMB constraints
obtained for a Planck-type experiment with 65\% sky coverage (Masahiro
Takada, private communication). The temperature and polarization power
spectra and the cross spectrum were computed using CMBFAST version
4.5.1. The Fisher matrix was initially calculated for 9 parameters:
$\omega_{\rm cdm}$, $\omega_{b}$, $\Omega_{DE}$, $w_{0}$, $w_{a}$,
$n$ (primordial power spectrum index), $A_{s}$ (amplitude of initial
scalar fluctuations), $\alpha$ (primordial running index), and $\tau$
(optical depth at recombination). This 9-dimensional matrix was then
projected into our 4-dimensional parameter space and the new matrix
was added to the Fisher matrix given by equation~(\ref{eq:fishPSV}).
Two tables corresponding to $(S/N)_{\rm{min}}=5$ and
$(S/N)_{\rm{min}}=10$ are shown below. 

We first note that the
constraints from the two experiments are remarkably similar.
LST measures $\sigma_{8}$ slightly better than SNAP, while the opposite
is true for $w_{a}$, the parameter describing the evolution of the dark energy
equation of state.  We expect this trend: due to its higher
detection threshold, LST takes more information from the steep
high-mass end of the mass 
function, where a variation in $\sigma_{8}$ is very acutely
felt. On the other hand, SNAP can see higher redshift
clusters and test a deeper survey volume than LST, so is more
sensitive to the evolution of the dark energy equation of state.

\begin{figure}[h!]  \centering
 \includegraphics[scale=0.8]{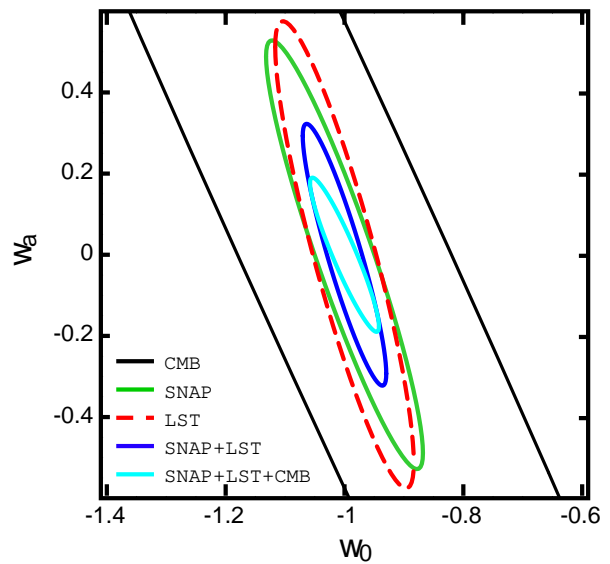}
\caption{Poisson+sample variance constraints on $w_{0}$ and $w_{a}$
when $(S/N)_{\rm{min}}=5$: From outermost to innermost, CMB is black,
SNAP is green, LST is red (dashed),
SNAP+LST is blue and finally the cyan ellipse is obtained by
combining SNAP+LST+CMB. The intrinsic ellipticities of galaxies are
the dominant noise here and we have marginalized over $\Omega_{m}$ and $\sigma_{8}$. }
\label{fig:ellipseLSTSNAP}
\end{figure}

The effect of sample variance on the dark-energy constraints is
minimal, $\le15\%$ for the SNAP survey and $<5\%$ for LST.
Adding CMB information brings down the constraints on all parameters,
but especially on $w_{a}$, by $\approx 50\%$ for both
instruments. This is further improved (by $40\%-50\%$) if we consider
LST and SNAP as independent experiments (assuming they probe
different survey volumes) and just sum their Fisher matrices. This can
be seen best in figure~\ref{fig:ellipseLSTSNAP}, where we have plotted
the constraints on the dark energy equation of state parameters for
both telescopes when $(S/N)_{\rm{min}}=5$.

Finally we note that dark energy constraints depend strongly on choice
of threshold $(S/N)_{\rm min}$.  Doubling the detection threshold to 10
degrades the cosmological constraints by a factor $\sim2$ for the
SNAP$+$CMB case, or factors of $\sim 3$ for LST$+$CMB.

We conclude that both instruments have similar performances and that,
at least for constraining dark energy, cluster counts are useful when
combined with other experiments. Our results are far less optimistic
than the predictions given by \cite{WK2}, but we shall defer
comparisons with other papers to the last section.

\begin{table}[h!]
   \begin{center}
        \begin{tabular}{|c|c|c|c|} \hline\hline
\emph{$\rm{(S/N)_{\!min}=5}$} &
\multicolumn{3}{c|}{\emph{SNAP}} \\ \hline \emph{\,}& \emph{Poisson} &
\emph{Poisson+SV}& \emph{Add CMB} \\ \hline\hline $\Omega_{m}$ & 0.005
& 0.006 & 0.005 \\ \hline $\sigma_{8}$ & 0.006 & 0.007 & 0.006 \\
\hline $w_{0}$ & 0.078 & 0.087 & 0.069 \\ \hline $w_{a}$ & 0.304 &
0.348 & 0.197 \\ \hline\hline \emph{\,}&
\multicolumn{3}{c|}{\emph{LST}} \\ \hline $\Omega_{m}$ & 0.004 & 0.005
& 0.004 \\ \hline $\sigma_{8}$ & 0.004 & 0.004 & 0.004 \\ \hline
$w_{0}$ & 0.076 & 0.077 & 0.062 \\ \hline $w_{a}$ & 0.373 & 0.380 &
0.182 \\ \hline\hline \emph{\,}& \multicolumn{3}{c|}{\emph{LST+SNAP}}
\\ \hline $\Omega_{m}$ & 0.003 & 0.003 & 0.003 \\ \hline $\sigma_{8}$
& 0.003 & 0.003 & 0.003 \\ \hline $w_{0}$ & 0.043 & 0.046 & 0.038 \\
\hline $w_{a}$ & 0.197 & 0.213 & 0.125 \\ \hline\hline
\end{tabular}
          \caption{1-$\sigma$ constraints obtained for $(S/N)_{min}=5$,
considering only the intrinsic ellipticity noise.}
          \label{SN5}
    \end{center}
\end{table}

\begin{table}[h!]
   \begin{center}
        \begin{tabular}{|c|c|c|c|} \hline\hline
\emph{$\rm{(S/N)_{\!min}=10}$} &
\multicolumn{3}{c|}{\emph{SNAP}} \\ \hline \emph{\,}& \emph{Poisson} &
\emph{Poisson+SV}& \emph{Add CMB} \\ \hline\hline $\Omega_{m}$ & 0.014
& 0.014 & 0.012 \\ \hline $\sigma_{8}$ & 0.013 & 0.013 & 0.011 \\
\hline $w_{0}$ & 0.203 & 0.204 & 0.147 \\ \hline $w_{a}$ & 0.915 &
0.940 & 0.397 \\ \hline\hline \emph{\,}&
\multicolumn{3}{c|}{\emph{LST}} \\ \hline $\Omega_{m}$ & 0.015 & 0.015
& 0.013 \\ \hline $\sigma_{8}$ & 0.011 & 0.011 & 0.008 \\ \hline
$w_{0}$ & 0.268 & 0.269 & 0.190 \\ \hline $w_{a}$ & 1.590 & 1.600 &
0.548 \\ \hline\hline \emph{\,}& \multicolumn{3}{c|}{\emph{LST+SNAP}}
\\ \hline $\Omega_{m}$ & 0.007 & 0.007 & 0.007 \\ \hline $\sigma_{8}$
& 0.006 & 0.006 & 0.005 \\ \hline $w_{0}$ & 0.121 & 0.123 & 0.092 \\
\hline $w_{a}$ & 0.644 & 0.661 & 0.268 \\ \hline\hline
\end{tabular}
          \caption{1-$\sigma$ constraints obtained for $(S/N)_{min}=10$,
considering only the intrinsic ellipticity noise.}
          \label{SN10}
    \end{center}
\end{table}

\section{Large scale structure projections}
\label{projection}

\par It has been anticipated that WL cluster measurements will be
substantially compromised by the so-called projection effects: the
lensing signal could be produced by any structures along the line of
sight, not just virialized clusters. This effect causes uncertainties
in cluster mass determinations and consequently in cluster abundancies
(e.g. \cite{D} for a comprehensive discussion of errors on mass
estimations and solutions to reduce them; see also
\cite{Hoe1,Hoe2,Ma05}). \cite{HS4} study numerically the efficiency of 
locating clusters in shear maps and conclude that about 15\% of the
most significant peaks detected in noiseless WL maps do not have a
collapsed halo with mass greater than $10^{13.5}M_{\odot}/h$ within a
3' aperture; also see the related work of \cite{Takada}.
\par The purpose of this section is to determine how important the
large scale projections are for cluster counting and parameter
constraints.  We shall first consider the lensing signal of clusters with
masses $\textit below$ the detection threshold computed in
\S\ref{snmax} as the $\textit only$ source of noise for the projected
mass measurement (i.e. we ignore completely the intrinsic shape noise of
galaxies). We shall also regard the signal of these smaller clusters
as wholly uncorrelated with the signal of detectable clusters employed
for the estimates in \S\ref{eonly}, presuming them to be at widely
separated redshifts. Given these assumptions, we would
like to establish the minimum detectable mass of a cluster at an
arbitrary redshift $z_{d}$. We follow the steps of section
\S\ref{snmax} to calculate the variance of the mass estimator \tmass,
replacing the intrinsic-ellipticity noise with the projection 
noise. We retain the weights derived in \S\ref{snmax} by optimizing
the $S/N$ for the intrinsic-ellipticity noise.
Although this is no longer an optimal
filter, it allows us to compare the shape noise and the projection
noise and to establish the redshift regime where each of them
dominates.

\subsection{The mass estimator}

When the noise of the small structures along the line of sight
dominates our measurement the contributions of different
source-redshift bins are correlated, so we need to calculate the
tangential shear power spectrum of these bins, as indicated by
~(\ref{eq:FullVar}). Since the convergence power spectrum is easier to
estimate than the tangential shear spectrum, we reexpress the mass
estimator in terms of convergence rather than shear:
\begin{equation} 
\tmass=\int d^{2}\theta\,\sum_{i}w_{\kappa}(\theta,z_{i})
\kappa(\theta,z_{i})\,, \label{eq:newM2}
\end{equation} 
where we have considered directly the case of continuous annuli.  The
new convergence weights are linked to the old shear weights in the
following manner (see for instance \cite{Schneider}):
\begin{equation}
w_{\kappa}(x)=2\int_{x}^{\infty}dy\,\frac{w_{\gamma}(y)}{y}
\,-w_{\gamma}(x)\label{eq:linkw}\,.
\end{equation}

As already said, we take the same shear weights that optimize the shot
noise filter and manipulating a little equation~(\ref{eq:linkw}) we
obtain for the convergence weights:
\begin{equation} 
w_{\kappa}(\theta,z)=\left\{\begin{array}{ll}
\frac{\mathcal{C}}{\frac{\sigma_{\gamma}^{2}}{n}}
\left[\kappa(\theta,z)-\bar{\kappa}(\theta_{lim},z)\right]\,,\;\;
\theta\leq\theta_{lim}\\ 0,\;\;\; \theta>\theta_{lim}\;,
\end{array}\right. \label{eq:weights}
\end{equation} 
with $\theta_{lim}$ defined in \S\ref{snmax} and the mean convergence
inside a radius $\theta$, $\bar{\kappa}(\theta,z)=\frac{2}{\theta^{2}}
\int_{0}^{\theta}dx\,x\,\kappa(x,z)$. Just like in \S\ref{snmax}, we
shall use the convergence of a canonical cluster with an NFW profile
density. $\mathcal{C}$ is the constant defined by
equation~(\ref{eq:constant}).  \par We write the estimator defined
by~(\ref{eq:newM2}) in Fourier space and take its variance:
\begin{equation} 
{\rm Var}(\tmass)=\frac{1}{(2\pi)^{2}}\sum_{i,j}
\int\,d^{2}l\,w_{\kappa}(\vec{l},z_{i})\, 
\mathcal{P}_{\rm{\kappa}}^{ij}(l)\,w_{\kappa}(\vec{l},z_{j})\,,
\end{equation} 
where we have used the definition of the convergence power spectrum:
$$\langle
\kappa(\vec{l},z_{i})\,\kappa(\vec{l'},z_{j})\rangle\equiv(2\pi)^{2}\,\mathcal{P}_{\kappa}^{ij}(l)\,\delta_{D}(\vec{l}-\vec{l'})\,,$$
and the Fourier transform of the convergence weight:
$$w_{\kappa}(\vec{l},z)=2\pi\int_{0}^{\infty}d\theta\,\theta\,w_{\kappa}(\theta,z)\,J_{0}(l\,\theta)\,.$$
The power spectrum for bins $i,j$ is: (see for instance
\cite{Bhuvnesh})
\begin{eqnarray}
& & \mathcal{P}_{\rm{\kappa}}^{ij}(l)=
\left(\frac{3}{2} \frac{H_{0}^2}{c^2}\Omega_{m}\right)^{2}
\,\int_{0}^{\infty}dz_{p}\left|\frac{d\chi}{dz_{p}}
\right|(1+z_{p})^{2} 
\nonumber\\
& & \hspace{1.1cm} \times \;W_{i}(z_{p})W_{j}(z_{p})
P_{\rm{\delta}}\left(\frac{l}{\chi(z_{p})},z_{p}\right) 
\nonumber 
\end{eqnarray}
$P_{\rm{\delta}}$ is the 3D matter power spectrum and
$\chi$ is the comoving distance.  The source weights are given by the
expression:

$$W_{i}(z)=\frac{\int_{z_{i}}^{z_{i+1}}\,dz_{s}\mathcal{P}(z_{s})\,Z(z_{s},z)}{\int_{z_{i}}^{z_{i+1}}\,dz_{s}\mathcal{P}(z_{s})}\,.$$
In the limit of infinite number of redshift bins this simplifies to:
$W_{i}(z)=Z(z_{i},z)\,.$ \par Combining all these ingredients and
using ~(\ref{eq:weights}), the projection noise is thus defined:
\begin{eqnarray} 
& & \hspace{-0.45cm} 
{\rm Var}(\tmass,z_{d}) 
=\mathcal{C}^{2}\!\left(\frac{3}{2}\frac{H_{0}^2}{c^2}\Omega_{m}\right)^{2}
\nonumber\\
& & \hspace{0.5cm}
\times \int d^{2}l\,\left\{ d\theta\,\theta
\left[\kappa_{\infty}(\theta,z_{d})-\bar{\kappa}_{\infty}
(\theta_{lim},z_{d})\right]J_{0}(l\theta)\right\}^{2}
\nonumber\\
& & \hspace{0.52cm}
\times\int_{0}^{\infty}dz_{p}\left|\frac{d\chi}{dz_{p}}\right|
(1+z_{p})^{2}P_{\rm{\delta}}\left(\frac{l}{\chi(z_{p})},z_{p}\right)
\nonumber\\
& & \hspace{0.55cm}
\times\left[\int_{0}^{\infty}dz_{s}\,Z(z_{s},z_{d})Z(z_{s},z_{p})
\mathcal{P}(z_{s})\right]^{2}\label{eq:newM3}
\end{eqnarray}
The nonlinear matter power spectrum is computed using the halo model
\cite{Rob}: it is the sum of a quasi-linear term and a halo term. The
quasi-linear term gives the power resulting from correlations of
distinct halos and dominates on large scales. The halo term describes
the correlations of particles within the same halo and dominates on
small scales.  At every $z_{p}$ (projection redshift),
$P_{\rm{\delta}}$ is estimated as the power of the virialized
structures with masses {\em smaller} than $M_{{\rm min}}(z)$ from
section \S\ref{snmax}.  Recall that halos above this mass are detected
as clusters and are hence part of the {\em signal}, so do not
contribute to the noise variance in the cluster-counting experiment.

\subsection{Projection Noise vs. Ellipticity Noise}
\label{proj results}

\par Equation~(\ref{eq:newM3}) represents the projection noise, caused
by the ``unseen'' small clusters that our featured surveys cannot
detect, but which contribute nonetheless to the whole lensing signal.
Solving again equation~(\ref{eq:Mmin}) with the variance given
by~(\ref{eq:newM3}) we find $M_{{\rm min}}^{\rm{proj}}(z)$ associated
with this noise.  In figure~\ref{fig:Mcrit5} we have plotted the
minimum detectable mass corresponding to the shape noise (solid line)
and to the projection noise (dotted line) for LST and SNAP when
$(S/N)_{\rm{min}}$ is 5. We note that projection noise exceeds
intrinsic-shape noise only for $z<0.07$ for LST, or slightly higher
($z<0.11$) for SNAP due to its lower shape-noise level.  If
$(S/N)_{\rm{min}}$=10, the crossover redshifts are 0.1 and 0.4,
respectively.

There are 2 factors that decide the function
$M_{\rm{min}}^{\rm{proj}}(z)$. Its shape is determined by the sources'
redshift distribution and its amplitude by the magnitude of the power
spectrum that we integrate in~(\ref{eq:newM3}). To understand why the projection noise is so high at small redshifts compared to the shape noise, we have to look at the way $S/N$ scales with $D_{d}$ (the lens-observer angular diameter distance) in each case. At small redshifts, the signal gets weak, because the shear gets weak. When the shape noise dominates, $S/N$ does not have a dependence on $D_{d}$ because the signal scales with $D_{d}$ the same way the noise does. This is not true for projection noise, where the noise and the signal scale differently, so $S/N \rightarrow 0$  as $z \rightarrow 0$ at fixed mass. A cluster
produces a maximum lensing signal if it lies halfway between the
source and the observer: $M_{\rm{min}}^{\rm{proj}}$ is minimal at the
redshifts obeying this condition. For LST, $\langle z\rangle$ is 1, so
the minimum occurs around 0.5. For SNAP it occurs at redshifts around
0.7, since $\langle z \rangle$ is 1.5 in this case. At higher
redshifts LST has fewer sources than SNAP, so there
$M_{\rm{min}}^{\rm{proj}}$ is higher for LST than for SNAP.

$M_{\rm{min}}^{\rm{proj}}$ is even more sensitive to the choice of the
detection threshold than $M_{\rm{min}}^{\rm{shape}}$. In the case of
shape noise, the variance of the mass estimator does not depend on
$(S/N)_{\rm{min}}$, as one can see from
equation~(\ref{eq:noise3}). $(S/N)_{\rm{min}}$ impacts
$M_{\rm{min}}^{\rm{shape}}$ only when we solve~(\ref{eq:Mmin}). In the
case of projection noise, the variance itself depends on
$(S/N)_{\rm{min}}$ through the power spectrum in
equation~(\ref{eq:newM3}).  For every projection redshift, we have
excluded from the total matter power spectrum the clusters with masses
bigger than $M_{\rm{min}}^{\rm{shape}}$ corresponding to that
redshift, because they are individually identifiable as clusters and
can be removed from the noise background. Therefore, the remaining
noise spectrum is stronger when 
$(S/N)_{\rm{min}}=10$ than when $(S/N)_{\rm{min}}=5$. In the same
vein, $M_{\rm{min}}^{\rm{proj}}$ is greater for LST than for SNAP,
because $M_{\rm{min}}^{\rm{shape}}$ is greater for LST than for SNAP.
The importance of $(S/N)_{\rm{min}}$ is further propagated when we
solve equation~(\ref{eq:Mmin}) for the variance given
by~(\ref{eq:newM3}). This is the reason why the difference between
$M_{\rm{min}}^{\rm{proj}}$ obtained for $(S/N)_{\rm{min}}=5$ and
$M_{\rm{min}}^{\rm{proj}}$ for $(S/N)_{\rm{min}}=10$ is greater than
the difference between $M_{\rm{min}}^{\rm{shape}}$ corresponding to
the same detection thresholds of 5 and 10.

In order to see the role of the projection noise in the cosmological
parameter constraints, we have added the shape-noise critical mass and
the projected-noise critical mass in quadrature and repeated the
calculations from \S\ref{eonly}. And just like in
\S\ref{eonly}, we took the variation with cosmology of the new total
critical mass into account.  Figure~\ref{fig:ncANs5} shows the number
of detectable clusters per unit redshift as a function of redshift
when we consider only the shape noise (solid line) and both the shape
and projection noises (dashed line). The total number of clusters does
not decrease significantly ($\le10\%$) when the projection noise is
included, if $(S/N)_{\rm{min}}=5$.  There is negligible change to the
derived cosmological constraints.

\begin{figure}[h!]  \centering
 \includegraphics[scale=0.8]{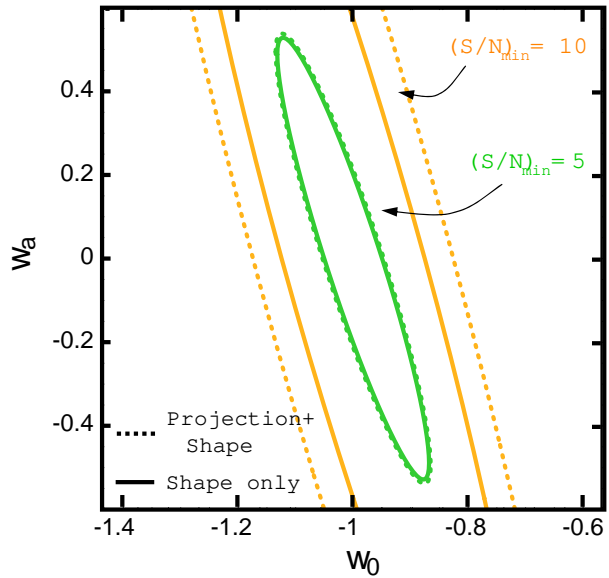}
\caption{SNAP constraints on $w_{0}$ and $w_{a}$: the orange (outer)
  solid ellipse
is for $(S/N)_{\rm{min}}=10$ and shape noise only, the green (inner) solid
ellipse is for $(S/N)_{\rm{min}}=5$ and shape noise only. The 
dotted ellipses are obtained for $(S/N)_{\rm{min}}=5$ and when we add
the projection noise to the shape noise. These are all Poisson+sample
variance errors and we have marginalized over $\Omega_{m}$ and $\sigma_{8}$. }\label{fig:ellipseSNAP}
\end{figure}

Raising the threshold to
$10\sigma$ can cut the number of detectable clusters more severely,
particularly for SNAP, where it eliminates 60\% of detections.
The effect upon derived parameter constraints remains small, however,
if we combine the cluster surveys with either the CMB or with each
other. 
Figure~\ref{fig:ellipseSNAP} shows the constraints on $w_{0}$ and
$w_{a}$ obtained for SNAP first for ellipticity noise only and then
when projections are included.

If the projection noise were to become a significant contributor to
the error budget, one could construct an estimator that maximizes the
signal-to-noise in the presence of both shape noise and projection
noise \cite{D,Ma05}.

\begin{table}[h!]

   \begin{center}
        \begin{tabular}{|c|c|c|c|} \hline\hline
\emph{$\rm{(S/N)_{\!min}=5}$} &
\multicolumn{3}{c|}{\emph{SNAP}} \\ \hline \emph{\,}& \emph{Poisson} &
\emph{Poisson+SV}& \emph{Add CMB} \\ \hline\hline $\Omega_{m}$ & 0.006
& 0.006 & 0.006 \\ \hline $\sigma_{8}$ & 0.006 & 0.007 & 0.006 \\
\hline $w_{0}$ & 0.082 & 0.090 & 0.072 \\ \hline $w_{a}$ & 0.315 &
0.355 & 0.203 \\ \hline\hline \emph{\,}&
\multicolumn{3}{c|}{\emph{LST}} \\ \hline $\Omega_{m}$ & 0.005 & 0.005
& 0.005 \\ \hline $\sigma_{8}$ & 0.004 & 0.004 & 0.004 \\ \hline
$w_{0}$ & 0.077 & 0.078 & 0.063 \\ \hline $w_{a}$ & 0.380 & 0.386 &
0.185 \\ \hline\hline \emph{\,}& \multicolumn{3}{c|}{\emph{LST+SNAP}}
\\ \hline $\Omega_{m}$ & 0.003 & 0.003 & 0.003 \\ \hline $\sigma_{8}$
& 0.003 & 0.003 & 0.003 \\ \hline $w_{0}$ & 0.044 & 0.047 & 0.038 \\
\hline $w_{a}$ & 0.201 & 0.216 & 0.126 \\ \hline\hline
\end{tabular}
          \caption{1-$\sigma$ constraints obtained for $(S/N)_{min}=5$ when we
add the projection noise to the intrinsic ellipticity noise.}
          \label{ProjNSN5}
    \end{center}
\end{table}

\section{Discussion and Conclusions}
\label{conc}

\par In this paper we have tried to determine how useful WL-detected
clusters are for constraining cosmology. We focus our attention
on the matter density parameter ($\Omega_{m}$), the power spectrum
normalization ($\sigma_{8}$) and the time-evolving dark energy
equation of state ($w_{0},w_{a}$), assuming a flat universe.

\par There are a few important features in our Fisher matrix analysis
that distinguish it from previous work.
We compute the Fisher
matrix using as an observable the directly measured quantity, the
filtered shear \tmass.  
Assuming some density profile for the detected lenses,
their measured shear can be converted to a virial mass. However, this
conversion depends on the cosmology. In different cosmologies,
different masses correspond to the same measured shear. Our Fisher
matrix takes this fact into account. We have also employed the shape
of the mass function, by keeping track of the \tmass\ of the objects in
bins rather than merely counting clusters above a threshold.
A third distinction of our analysis is the halo-model calculation of
the noise variance due to uncorrelated projections along the line of
sight.  As noted above, this is shown to cause little loss of
cosmological information in most circumstances.

When we apply our formalism to canonical ground (LST) and space (SNAP)
WL surveys, we find the two are nearly equivalent.  The lower
cluster-mass threshold afforded by the deeper sample of source
galaxies in space provides a 15-fold increase in the sky density of
detected clusters, which compensates for the smaller survey solid
angle that we assume for the SNAP survey.  Both surveys produce quite
interesting dark-energy constraints, and improve significantly when
combined with CMB constraints.  Combining the LST and SNAP surveys
results in constraints significantly stronger than either alone.

Comparison with previous forecasts for WL cluster surveys is
complicated by the extreme sensitivity to the assumed noise level and
detection threshold of the survey---this behavior is of course
generically true of cluster-counting experiments.  

\par Our results are less encouraging than some other forecasts in the
literature:  for a detection threshold of 5 and considering only the
intrinsic ellipticities of galaxies, LST could detect about 21500
clusters. The calculated LST errors on cosmological parameters are:
$\Delta \Omega_{m}=0.005$, $\Delta \sigma_{8}=0.004$,
$\Delta w_{0}=0.08$, $\Delta w_{a}=0.38$, if we take into account both
the Poisson and sample variance noises. SNAP yields rather similar
constraints.

These values obtained for the dark energy equation of state are a few
times higher than found by \cite{WK2}: $\Delta w_{0}=0.05$ and
$\Delta w_{a}=0.09$. These authors analyze an LST-type experiment, using a
Gaussian filter for the shear signal, as proposed by
\cite{Takada}. Such a difference in our results is partly (but not
completely) explained by the assumed parameters of their survey. The angular
concentration of galaxies is 65 galaxies/$\rm{arcmin}^{2}$, the survey
area is 18000$\rm{deg}^{2}$ and $\sigma_{\gamma}$ is 0.15, 
more optimistic than our corresponding values of 30
galaxies/$\rm{arcmin}^{2}$, 15000$\rm{deg}^{2}$ and 0.3.  Additionally,
they take $(S/N)_{\rm{min}}=4.5$. They estimate
about 200000 clusters are detectable, $10\times$ higher than our
nominal estimate.  If we assume their input values, 
our optimal filter yields about 700000 clusters, demonstrating the
advantage of an optimal filter.  
\par The last issue that we discussed is that of the projection
contamination, considered the sword of Damocles for WL measurements.
From the beginning
we make a crucial approximation: we identified the projected mass of
clusters with their virial mass.  This need not be true on a
cluster-by-cluster basis, just in the sense that the mass functions
have similar amplitude and dependence upon cosmological parameters.
With this assumption---which needs to be verified numerically---about
the effect of {\em local} structure on projected-mass estimates, we can
next treat the projection of {\it unrelated} structures along the line
of sight as a noise source, not a bias.

\par WL cluster measurements are subject to two major sources of
noise: the intrinsic ellipticities of galaxies (shape noise) and these
large scale projections.  We find that, even when we use an
overdensity-detection filter optimized for pure shape noise, 
that the projection noise is dominant only for lenses at
$z_d\lesssim0.1$, and has little effect upon the derived dark-energy
constraints in most applications of our canonical ground (LST) and
space (SNAP) WL cluster surveys.

Let us stress again that all these numbers are
extremely sensitive to the detection threshold and the noise levels
that we impose. If $(S/N)_{\rm{min}}=10$, the number of detectable
clusters goes down by a factor of 10 in the case of shape noise only
and by more than a factor of 18 when we also account for
projections. Then the errors on the dark energy equation of state
parameters are as much as tripled.
If the noise has
Gaussian statistics, as we would expect for shape noise, then the
$5\sigma$ threshold will suffice to keep false positives to an
unimportant level.  The projection noise will not, however, be
Gaussian, so further investigation is required to determine whether
the $5\sigma$ threshold offers enough suppression.  The numerical
study of \cite{HS4} suggests this is the case, as their $4.5\sigma$
threshold results in a false-positive contamination of only 25\%.
Recall that this work considers any object not associated with a 
virialized halo to be a false positive, whereas we assert that
counting unvirialized objects is a valid cosmological test, so this
25\% ``contamination'' still carries cosmological information.

We conclude, therefore, that counting of WL-detected clusters can be a
powerful constraint on cosmology with either space- or ground-based
surveys.  While projection effects make it difficult to establish a
one-to-one correspondence between WL-derived masses and virial masses,
there is no real need to make such a correspondence in order to infer
cosmological parameters from WL data.  Indeed it is precisely the
ability to go directly from $N$-body simulations to WL-derived masses
that makes WL cluster counting an attractive alternative to optical,
X-ray, or SZ cluster counting.  Our Fisher analysis incorporates the
cosmological dependence of the conversion from shear to mass units,
plus the effect of sample variance and projection noise, and we show
that none of these are barriers to strong dark-energy constraints.

\section*{Acknowledgments}
LM is supported by grant AST-0236702 from the
National Science Foundation.  GMB acknowledges additional support from
Department of Energy grant DOE-DE-FG02-95ER40893 and NASA BEFS-04-0014-0018.
We are very grateful to Robert Smith for letting us use his halo model matter
power spectrum code and to Jacek Guzik for carefully reading this
manuscript. We thank Bhuvnesh Jain, Masahiro Takada, Licia Verde and
Wayne Hu for their gracious assistance.

\end{document}